\documentclass[12pt]{iopart} 

\usepackage{iopams,amsfonts,amssymb,amsthm} 

\pdfoutput=1
\usepackage{etex}
\usepackage{cite}
\usepackage{latexsym}
\usepackage{rawfonts}
\usepackage{graphicx}
\usepackage[usenames,dvipsnames,svgnames]{xcolor}
\usepackage{bbm}
\usepackage{latexsym}
\usepackage{multirow}
\usepackage{rotating}
\usepackage{lscape}
\usepackage{graphicx} 
\usepackage{subfigure}
\usepackage{xcolor}
\usepackage{fancybox}
\usepackage{ifmtarg}

\input prepictex
\input pictex
\input postpictex

\usepackage{times}

\usepackage[font=footnotesize,labelfont=sf]{caption}

\def\2;{\;\;}

\def\IntN{{\mathbb Z}}

\def\Ref#1{(\ref{#1})}

\def\C#1{{\mathcal #1}}

\def\Sfrac#1#2{\hbox{\large $\frac{#1}{#2}$}}
\def\sfrac#1#2{\hbox{\nor $\frac{#1}{#2}$}}

\def\LB{\left(}         \def\RB{\right)}

\def\nor{\normalsize}
\def\fns{\scriptsize}

\def\lvv{\hbox{\LARGE$|$}}

\definecolor{blue}{rgb}{0,0.18,0.39}
\definecolor{RoyalBlue}{rgb}{0,0.2,0.7}

\def\axes#1#2#3#4#5#6#7{
\setplotarea x from #7 to #5, y from #2 to #6
\setplotarea x from #1 to #5, y from #2 to #6
\axis left shiftedto x=#3 
        ticks
        withvalues #6 #2 /
        at  #6 #2 /
 /
\axis bottom shiftedto y=#4
        ticks
        withvalues #1 #5 /
        at  #1 #5 /
/
\put {\footnotesize$\bullet$} at #3 #4
}

\def\axesnolabels#1#2#3#4#5#6#7{
\setplotarea x from #7 to #5, y from #2 to #6
\setplotarea x from #1 to #5, y from #2 to #6
\axis left shiftedto x=#3 
        ticks
        at  #6 #2 /
 /
\axis bottom shiftedto y=#4
        ticks
        at  #1 #5 /
/
\put {\footnotesize$\bullet$} at #3 #4
}

\definecolor{color35}{cmyk}{0,0.32,0.74,0.875}
\definecolor{color39}{cmyk}{0,0.32,0.74,0.975}
\definecolor{color40}{cmyk}{0,0.32,0.74,1}

\definecolor{B03}{cmyk}{0,0.12,0.22,0.12}
\definecolor{B04}{cmyk}{0,0.14,0.24,0.14}
\definecolor{B05}{cmyk}{0,0.16,0.26,0.16}
\definecolor{B06}{cmyk}{0,0.19,0.28,0.19}
\definecolor{B07}{cmyk}{0,0.22,0.30,0.22}
\definecolor{B08}{cmyk}{0,0.25,0.32,0.25}
\definecolor{B09}{cmyk}{0,0.28,0.34,0.28}
\definecolor{B10}{cmyk}{0,0.31,0.36,0.31}
\definecolor{B11}{cmyk}{0,0.34,0.38,0.34}
\definecolor{B12}{cmyk}{0,0.37,0.40,0.37}
\definecolor{B13}{cmyk}{0,0.4,0.42,0.40}
\definecolor{B14}{cmyk}{0,0.43,0.44,0.43}
\definecolor{B15}{cmyk}{0,0.46,0.46,0.46}
\definecolor{B16}{cmyk}{0,0.49,0.48,0.49}
\definecolor{B17}{cmyk}{0,0.52,0.50,0.52}
\definecolor{B18}{cmyk}{0,0.55,0.52,0.54}
\definecolor{B19}{cmyk}{0,0.58,0.54,0.56}
\definecolor{B20}{cmyk}{0,0.61,0.56,0.58}
\definecolor{B21}{cmyk}{0,0.64,0.58,0.60}
\definecolor{B22}{cmyk}{0,0.69,0.61,0.62}
\definecolor{B23}{cmyk}{0,0.72,0.64,0.64}
\definecolor{B24}{cmyk}{0,0.75,0.67,0.66}
\definecolor{B25}{cmyk}{0,0.78,0.70,0.68}
\definecolor{B26}{cmyk}{0,0.81,0.73,0.70}
\definecolor{B27}{cmyk}{0,0.84,0.76,0.70}
\definecolor{B28}{cmyk}{0,0.87,0.79,0.68}
\definecolor{B29}{cmyk}{0,0.90,0.82,0.65}
\definecolor{B30}{cmyk}{0,0.93,0.85,0.60}
\definecolor{B31}{cmyk}{0,0.96,0.88,0.50}
\definecolor{B32}{cmyk}{0,0.99,0.91,0.30}
\definecolor{B33}{cmyk}{0,1,0.94,0.30}
\definecolor{B34}{cmyk}{0,1,0.97,0.30}
\definecolor{B35}{cmyk}{0.4,1,0.8,0.30}
\definecolor{B36}{cmyk}{0.6,1,0.6,0.30}
\definecolor{B37}{cmyk}{1,1,0.4,0.20}
\definecolor{B38}{cmyk}{1,1,0.2,0.15}
\definecolor{B39}{cmyk}{1,1,0,0.10}
\definecolor{B40}{cmyk}{1,1,0,0}

\begin{document}

\title{Osmotic Pressure of Confined Square Lattice Self-Avoiding Walks}
\author{
F Gassoumov$^1$\footnote[1]{\texttt{fgassoum@yorku.ca}}
\&
EJ Janse van Rensburg$^1$\footnote[2]{\texttt{rensburg@yorku.ca}}
}

\address{\sf$^1$Department of Mathematics and Statistics, 
York University, Toronto, Ontario M3J~1P3, Canada\\}

\date{\today} 

\begin{abstract} \sf
Flory-Huggins theory is a mean field theory for  modelling the free energy
of dense polymer solutions and polymer melts.  In this paper we use
Flory-Huggins theory as a model of a dense two dimensional self-avoiding 
walk compressed in a square in the square lattice.   The theory describes the 
free energy of the walk well, and we estimate the Flory interaction
parameter of the walk to be $\chi_{saw} = 0.32(1)$.
\end{abstract}

\maketitle

\section{Introduction}

Flory-Huggins theory \cite{Flory} is used widely in the modelling of polymer 
solutions and polymer melts.  The basic component in this theory is the 
mean field quantification of the entropy of a polymer.  If a single chain 
is considered, then the entropy per unit volume $V$ is estimated by
\begin{equation}
- S_{site} (\phi)= \Sfrac{\phi}{N} \log \Sfrac{\phi}{N} + (1-\phi)\log(1-\phi)
\label{eqn1}  
\end{equation}
where $\phi = \sfrac{N}{V}$ is the volume fraction (or concentration) of monomers 
in a chain of length $N$ (degree of polymerization) confined in a space 
of volume $V$.   
In Flory-Huggins theory the full entropy  $S_{site}(\phi)$ is modified to the 
\textit{entropy of mixing} $S_{mix}$ \cite{Flory42} per site, which is the difference between 
$S_{site}(\phi)$ and the weighted average of the entropy of pure solvent 
$S_{site}(0)$, and pure polymer $S_{site}(1)$:
\begin{equation}
\fl \hspace{0cm}
- S_{mix} = - S_{site}(\phi) + (1-\phi) S_{site}(0) + \phi\, S_{site}(1)
= \Sfrac{\phi}{N} \log \phi + (1-\phi)\log (1-\phi) .
\end{equation}
Notice the cancellation in $S_{mix}$ of the term linear in $\phi$.

The \textit{free energy of mixing per site} $F_{mix}$ in Flory-Huggins theory is obtained
by adding to ${-}S_{mix}$ the energy of mixing $\sfrac{1}{T}E_{mix}$ 
per site, where $T$ is the temperature (see, for example, reference
\cite{deG79}).  This includes the energy contribution per site for 
monomer-solvent interactions and it is given by
$E_{MS} = T\chi_{MS}\, \phi(1-\phi)$.  The parameter $\chi_{MS}$ is assumed to be
a constant.  Similarly, there are terms accounting
for solvent-solvent and monomer-monomer interactions with parameters
$\chi_{SS}$ and $\chi_{MM}$, but when $E_{mix}$ is
calculated then these combine into a single parameter $\chi = \chi_{MS} -
\sfrac{1}{2}(\chi_{MM}+\chi_{SS})$ so that
\begin{equation}
E_{mix} = T\, \chi\, \phi(1-\phi).
\label{eqn3}   
\end{equation}
The total energy in the mean field
should be given by $E_{tot} = E_{mix}  + [\hbox{\textit{terms linear in $\phi$}}]$,
where the linear terms are contributions of individual monomers or solvent
molecules and are of the form $T\,\gamma_m\, \phi + T\,\gamma_s\, (1-\phi)$. The coefficients
$\gamma_m$ and $\gamma_s$ account for the changes in $E_{tot}$ if a single
monomer or solvent molecule is added or removed.  Equation \Ref{eqn3} also
ignores three body and higher order contributions which should be present in
the theory, for example, terms of the form $T\chi_3\, \phi^2(1-\phi)$.

The parameter $\chi$ in equation \Ref{eqn3}
is the dimensionless \textit{Flory interaction parameter} and it is a function
of temperature and pressure.   In this presentation, $\chi$ is a measure of
(repulsive) interactions between solvent molecules and monomers. In general 
$\chi$ is positive, and increasing with $T$ \cite{deG79} (but it can be negative).  
Low values of $\chi$ are indicative of good solvents (and if $\chi=0$ then the 
solution is said to be ``athermal").  When $\chi=0.5$ then the solvent is 
said to be marginal, 
and for $\chi>0.5$ the solvent is poor and may induce polymer collapse
\cite{deG75,S75} from a coil to a globule phase.

\begin{figure}[t]
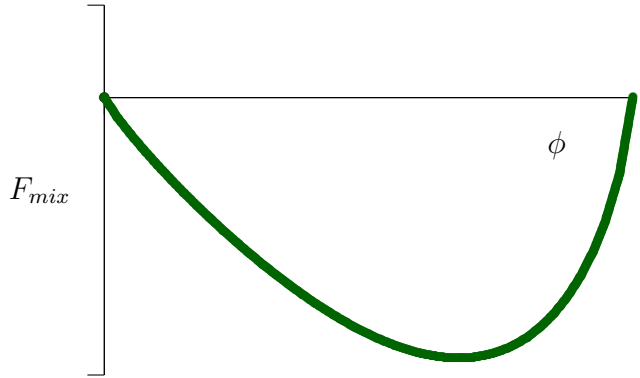

\input Figures/fig01.tex
\caption{The Flory-Huggins free energy of mixing $F_{mix}$ 
(see equation \Ref{eqn4}) plotted 
against $\phi$ with $N=10$ and $\chi=\frac{1}{2}$.}
\label{fig01}  
\end{figure}

The \textit{mean field free energy of mixing} $F_{mix}$ (per unit volume) is given by
\begin{equation}
\fl \hspace{0.5cm}
\Sfrac{1}{T}\, F_{mix} = \Sfrac{1}{T} \, E_{mix} - S_{mix}
= \Sfrac{\phi}{N} \log \phi + (1-\phi)\log(1-\phi) + \chi\, \phi(1-\phi) ,
\label{eqn4}   
\end{equation}
where linear terms are left away;
see, for example, equation III.7 in reference \cite{deG79}.  The general
shape of $F_{mix}$ is shown in figure \ref{fig01}.  If $\phi$ is small,
then this may be expanded to obtain
\begin{equation}
\Sfrac{1}{T}\, F_{mix} = \Sfrac{\phi}{N} \log \phi + \sfrac{1}{2} (1-2\chi)\, \phi^2 
+ \sfrac{1}{6} \phi^3 + \cdots ,
\label{eqn5}  
\end{equation}
and this shows that the (Edwards) \textit{excluded volume parameter} is
given by $\upsilon = 1-2\chi$ \cite{Edwards76}.  If $\upsilon=0$
(or $\chi=\sfrac{1}{2}$) then the polymer is in $\theta$-conditions and the first correction 
to $\sfrac{\phi}{N} \log \phi$ in $F_{mix}$ is the third order term in $\phi$.
If $\upsilon<0$ then the polymer is in a poor solvent, and if $\upsilon>0$ it is
in a good solvent.

In the mean field
presentation above $\chi$ is a parameter of Flory-Huggins theory, and so
is assumed to be independent of $\phi$ (but a function of temperature
$T$ and pressure $P$).  However, the mean field approach neglects correlations between
monomers in the derivation of equation \Ref{eqn1},  and so will not
be a good approximation if these correlations are essential.  A modification
of the expressions above is to assume that $\chi$ has dependence on
concentration for concentrated and semi-dilute solutions --  this was already
observed in the literature, see for example references \cite{CP14,GSC89,KS90} (see 
also the comments in reference \cite{deG79} in section III.1).    Generally,
Flory-Huggins theory is a mean field approximation, and will in some cases
not describe a model well.

There is a vast literature in polymer physics and chemistry devoted to
the measurement and calculation of $\chi$.  These studies use a variety of
experimental techniques in determining the Flory-Huggins parameters
for polymer blends and solutions \cite{LKB97,MV06,M07}, and, 
in reference \cite{KS90}, it is noted that \textit{"With regard to the variation 
of $\chi$ with $\phi$, greatly differing results are reported in 
the literature"}.  The values of $\chi$ for various blends are used 
as a measure of the miscibility of polymer melts \cite{LKB97} 
and solubility of polymer solutions,
with small values of $\chi$ indicating the compatibility of melts or
increased solubility of a polymer.

Concentration dependence of $\chi$ in a polymer-solvent system would
indicate that the polymer has varying degrees of solubility at different
concentrations.  For example, a small value of $\chi$ at low concentration
will indicate that the polymer dissolves easily, but an increasing value of
$\chi$ with rising concentration may suggest that the solvent
quality is deteriorating, even to the point that the polymer will precipitate
from the solvent at a critical concentration $\phi_c$.  This event is 
akin to polymer collapse \cite{deG75,S75} seen in poor solvents, but here it is
driven by the concentration of polymers.

\begin{figure}[t]
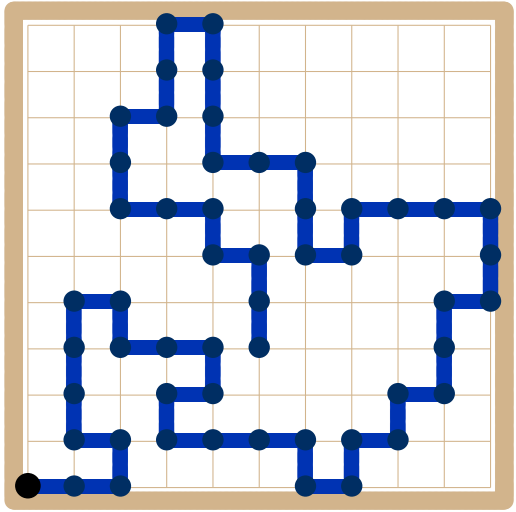

\input Figures/fig02.tex
\caption{A compressed self-avoiding walk confined to a square
of side length $L$ measured as the number of lattice sites along
the square (so that the square contains $L^2$ lattice sites).  In
this figure $L=11$.  The walk starts at the left-most, bottom-most
vertex in the square, and is confined to stay within
the square as it steps in the lattice.  The concentration of the
vertices in the compressed walk is given by $\phi=\frac{n+1}{L^2}$
where $n$ is the length of the walk. Vertices in the square
which are not occupied by the walk are the solvent molecules in this
model.}
\label{fig02}  
\end{figure}

These remarks raise the question of the applicability of Flory-Huggins
theory to theoretical models of dense polymer systems.  Below we
consider the usefullness of Flory-Huggins mean field expressions to
model the free energy and osmotic pressure of a self-avoiding walk
model of a compressed polymer.  We find that the data fit the mean
field expressions, with minor modification, quite well, and the model
may be a suitable framework for analysing compressed self-avoiding 
walks in the dense phase.

\subsection{Compressed self-avoiding walks}

A dense linear polymer in two dimensions is modelled by
a confined self-avoiding walk.  We confine
the walk to a square of dimensions $L\times L$ in $\IntN^2$ and
our aim is to model the free energy in terms of the Flory-Huggins 
formulation of the free energy.  Along the way we shall 
estimate the Flory interaction parameter $\chi$.  

The model is illustrated 
in figure \ref{fig02}, and the linear polymer is represented as a 
self-avoiding walk confined to a square of fixed side length $L$ (measured
as the number of sites along the side, so that the square contains
$L^2$ lattice sites).  The area or volume of the square is $L^2$.

The  walk of length $n$ steps (and $N=n+1$ vertices) starts in a corner 
of the square, and the concentration of the vertices
in the walk is given by $\phi = \sfrac{n+1}{L^2}$.  By  tuning the length of
the walk $n$, or the side length $L$ of the confining square, the 
concentration of the walk can be manipulated and its properties
examined as a function of $\phi$.  Unoccupied vertices in the square represent 
solvent molecules in the model, and their concentration is given by
$\phi_S = 1-\phi$.

Notice that an unconfined self-avoiding walk in the square lattice of
length $n$ has linear dimension $O(n^\nu)$ where $\nu$ is the metric
exponent.  It has Flory value $\nu=0.75$ in 2 dimensions \cite{F69} and this
is also its exact value \cite{N82,N84A}.  The concentration of vertices in the walk
in its convex hull is of order $O(\sfrac{n}{n^{2\nu}}) = O(n^{-0.5})$
which approaches $0$ as $n\to\infty$.  This shows that in the
$n\to\infty$ limit the concentration of vertices in an unconfined self-avoiding
walk is $\phi=0$.  In the context of the model in figure \ref{fig02} 
the limit as $\phi\to 0^+$ may be considered the (unconstrained) self-avoiding 
walk limit.  For non-zero concentrations ($\phi>0$) the walk is still 
self-avoiding, but it is constrained by the confining square, and we call it
a \textit{compressed walk}.

The configurational entropy of the compressed walk can be calculated from
the number of compressed walks of length $n$ in $\IntN^2$ and confined
to the square. If there are $c_{n,L}$ compressed walks from the origin at the bottom-most, 
left-most, corner of the $L\times L$ square, of length $n$ (see figure \ref{fig02}),
then the finite size (extensive) free energy of the model is 
\begin{equation}
F_{tot}(V) = - \log c_{n,L} .
\label{eqn7}   
\end{equation}
Notice that $F_{tot}(V)$ is both a function of $n$ and the area $V=L^2$
of the confining square and so is a function of $\phi$.

\begin{figure}[t!]
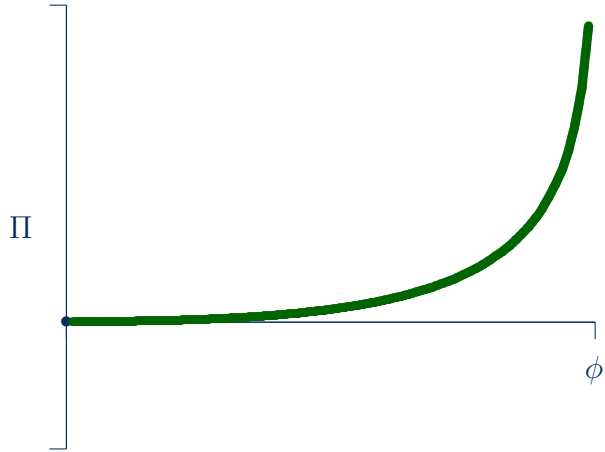

\input Figures/fig03.tex
\caption{A schematic diagram of the osmotic pressure calculated from the 
Flory-Huggins mean field field expression for the free energy of a 
compressed walk.  At low concentrations the pressure may be 
negative if $\chi>\frac{1}{2}$, but it is positive and increasing in a good solvent.}
\label{fig03}  
\end{figure}

The free energy per site (or unit area) is similarly a function of $\phi$ and is given by
\begin{equation}
F_V (\phi) = \Sfrac{1}{V}\, F_{tot}(V) = - \Sfrac{1}{V} \log c_{n,L} .
\label{eqn8A}  
\end{equation}
Mean field expressions for $F_V(\phi)$ is given by the expressions 
for $F_{mix}$ in equation \Ref{eqn4}, with the qualification that there may be
linear and higher order terms in $\phi$ missing in the mean field formula,
as observed earlier.

The free energy per unit length of the compressed 
walk is given by
\begin{equation}
f_t(\phi) = - \Sfrac{1}{n+1} \log c_{n,L}  = \Sfrac{1}{\phi}\, F_V(\phi) .
\label{eqn8B}  
\end{equation}
The self-avoiding walk limit is obtained when $L\to\infty$ and so when
$\phi\to 0^+$.   If $\phi=1$, then $f_t(1)$ is the free energy per unit
length of a Hamiltonian walk with one end-point at the bottom left
corner of a square.   The limit $\lim_{L\to\infty} f_t(1)$ should be
equal to $\kappa_H$, the square lattice \textit{connective constant} of Hamiltonian
walks.

\subsection{The osmotic pressure of a compressed self-avoiding walk}

The osmotic pressure of a macromolecule in solution is defined as the
variation of the free energy if the volume of solvent is changed while
the number of monomers $n$ is kept fixed.  In other words, in the context
here, by
\begin{equation}
\Sfrac{1}{T}\, \Pi = - \Sfrac{\Delta F_{tot}(V)}{\Delta V}\lvv_{n}
\label{eqn9}   
\end{equation}
where $F_{tot}(V)$ is the total free energy in equation \Ref{eqn7}
of a walk confined in a volume of size $V$.  In lattice units $T=1$, so that
the explicit dependence on $T$ is dropped in what follows.

\begin{figure}[t]
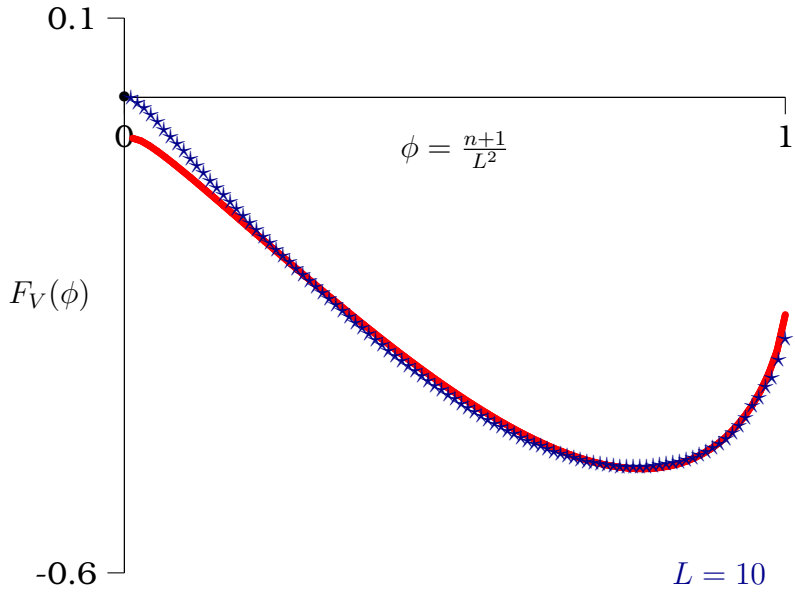

\input Figures/fig04.tex  
\caption{The estimated free energy per unit area $F_V(\phi)$ obtained from
flatPERM simulations for $L=10$ (denoted by the stars ($\star$)).  The free 
energy decreases quickly for small concentrations, goes through a minimum 
and then increases as the walk grows longer and loses entropy
when it is compressed by the confining square.  The curve underlying the data points
is the least squares fit to the data by equation \Ref{eqn15A}.  The fit is good,
except for the deviation at low concentration due to the divergence in the
term $\frac{1}{V}\log \phi$.}
\label{fig04}  
\end{figure}

The concentration of vertices along the walk of length $n$ (and with $n+1$ 
vertices) is given by $\phi = \sfrac{n+1}{V}$ so that $n+1 = \phi V$, 
and a change of variables to $\phi$ in equation \Ref{eqn9}  gives
\begin{equation}
\Pi = \phi^2\; \Sfrac{\partial}{\partial \phi} \LB \sfrac{1}{n+1} \,F_{tot}(V) \RB
=  \phi^2\; \Sfrac{\partial}{\partial \phi} \LB \sfrac{1}{\phi} \,F_V(\phi) \RB,
\label{eqn10}   
\end{equation}
in terms of $F_V(\phi)$.  Using equations \Ref{eqn7} and \Ref{eqn8B},
\begin{equation}
\Pi  =  \phi^2\; \Sfrac{\partial}{\partial \phi}  f_t(\phi) .
\label{eqn10A}   
\end{equation}

The free energy of mixing $F_{mix}$ (equation \Ref{eqn4}) is the mean
field expression, up to linear terms, of $F_V(\phi)$,
since both are free energy per unit area.  Thus, in terms of 
$F_{mix}$, the osmotic pressure has mean field estimate
\begin{equation}
\Pi = \phi^2 \; \Sfrac{\partial}{\partial \phi} \LB \sfrac{1}{\phi} \,F_{mix} \RB 
= \phi^2 \; \Sfrac{\partial}{\partial \phi} \LB \sfrac{1}{\phi} \,F_V \RB 
\label{eqn14}   
\end{equation}
in particular since all linear terms of $F_V$ in $\phi$ vanishes
if $F_{mix}$ is divided by $\phi$ and its derivative is taken.

In terms of the Flory-Huggins
free energy in equation \Ref{eqn4}, one should expect that
\begin{equation}
F_V(\phi) \approx a_1 \,\phi + \Sfrac{\phi}{n+1}\, \log \phi + (1-\phi)\log(1-\phi)
+ \chi\,\phi(1-\phi),
\label{eqn15}   
\end{equation}
where $a_1\phi$ is an additional linear term.  Expanding this in $\phi$ and 
collecting terms show that the linear dependence of $F_V(\phi)$ on
$\phi$ is $(a_1 + \chi - 1)\, \phi$.

Keeping $V$ fixed and taking the derivative to $\phi$ gives from equation \Ref{eqn10A}:
\begin{equation}
 \Pi = \phi^2\; \Sfrac{\partial}{\partial \phi} \LB \sfrac{1}{\phi} F_V (\phi) \RB
= \Sfrac{1}{V}  - \log(1-\phi) - \phi - \chi\,\phi^2  
\label{eqn13}  
\end{equation}
where we noted that $\sfrac{\phi}{n+1} = \sfrac{1}{V}$,  where $V=L^2$,
and where there is no dependence on the coefficient $a_1$ in equation
\Ref{eqn15}.  Taking $\phi\to0^+$ gives $\Pi\to\sfrac{1}{V}$, which is unphysical.
In the thermodynamic limit $V\to\infty$, and so the term $\sfrac{1}{V}$ can be
ignored.  See also the comments in reference \cite{deG79}, chapter III.1.3 on
the prediction that $\Pi \simeq \sfrac{1}{V}$ as $\phi \to 0^+$.  Later, in
our analysis in section 2.2, we will assume that we are sufficiently close to the
thermodynamic limit to ignore this term, and so leave it out of our model (see equation
\Ref{eqn24}).

For small $\phi$, $\Pi  = \sfrac{1}{V} 
+ (\sfrac{1}{2}-\chi)\,\phi^2 + \sfrac{1}{3}\,\phi^3 + \cdots$.
That is, if $\chi<\sfrac{1}{2}$ then $\Pi(\phi)>0$ for $\phi>0$.  On the other hand,
if $\chi>\sfrac{1}{2}$, then $\Pi(\phi) <0$ for $\phi$ small and $V$ large.  With increasing
$\phi$ the osmotic pressure becomes positive.  These approximations have 
to break down as $\phi\to 1^-$ because of the divergence of  the logarithmic 
term, so equation \Ref{eqn13} should be only valid in the dilute and semi-dilute regimes.

\begin{figure}[t]
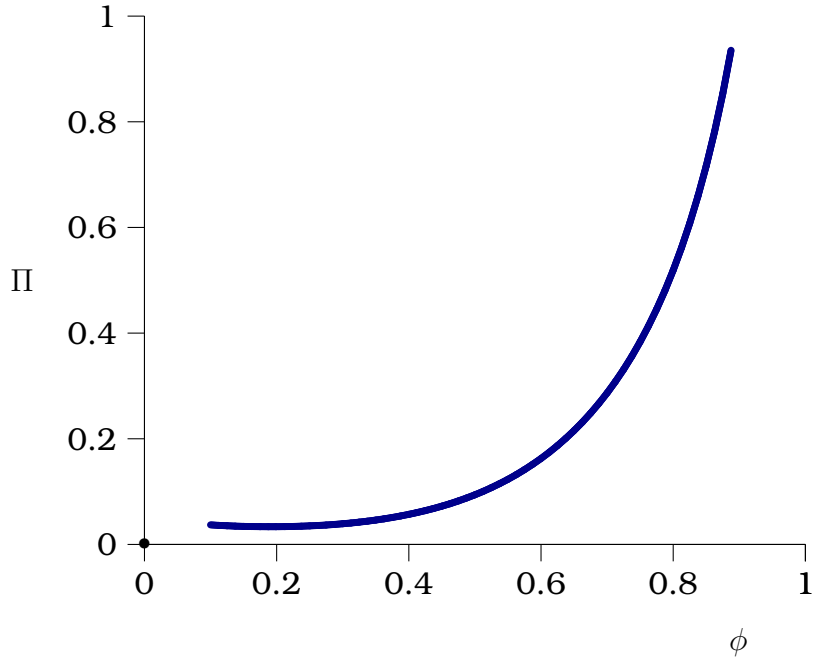

\input Figures/fig05.tex
\caption{A schematic diagram of the
osmotic pressure $\Pi$ of a square lattice self-avoiding walk in a 
$10\times 10$ square as a function of the concentration $\phi$.  
The pressure is positive for all $\phi>0$ and increases steadily with the
concentration $\phi$ (see equation \Ref{eqn19}).}
\label{fig05}  
\end{figure}

The term $\sfrac{1}{V}$ in equation \Ref{eqn13} arises from the contribution
$\Sfrac{\phi}{n+1}\, \log \phi$ in equation \Ref{eqn15} which, for large $n$, is small, 
and similarly, for small values of $\phi$, is small as well.  However,
there are contributions from this term for intermediate values of
$\phi$, and so we keep it in our model for the function $F_V(\phi)$
when $L$ is finite and fixed:
\begin{equation}
F_V(\phi) \approx a_0 \,\phi + \Sfrac{1}{V} \log \phi 
+ (1-\phi)\log(1-\phi) - \chi_L\,\phi^2,
\label{eqn15A}   
\end{equation}
where the linear term has coefficient $a_0 = (a_1 + \chi_L)$, and this
is a function of $L$.  We also indicated that the Flory Interaction parameter
$\chi_L$ may be a function of $L$.  The osmotic pressure is given by
\begin{equation}
 \Pi =  \phi^2\; \Sfrac{\partial}{\partial \phi} \LB \sfrac{1}{\phi} F_V (\phi) \RB
=  \Sfrac{1}{V} - \log(1-\phi) - \phi - \chi_L\,\phi^2  
\label{eqn13A}  
\end{equation}
where the first term $\sfrac{1}{V}$ should be ignored for large values of $L$,
and is also unphysical since $\Pi(\phi)$ should approach zero as $\phi\to 0^+$.

\begin{figure}[t]
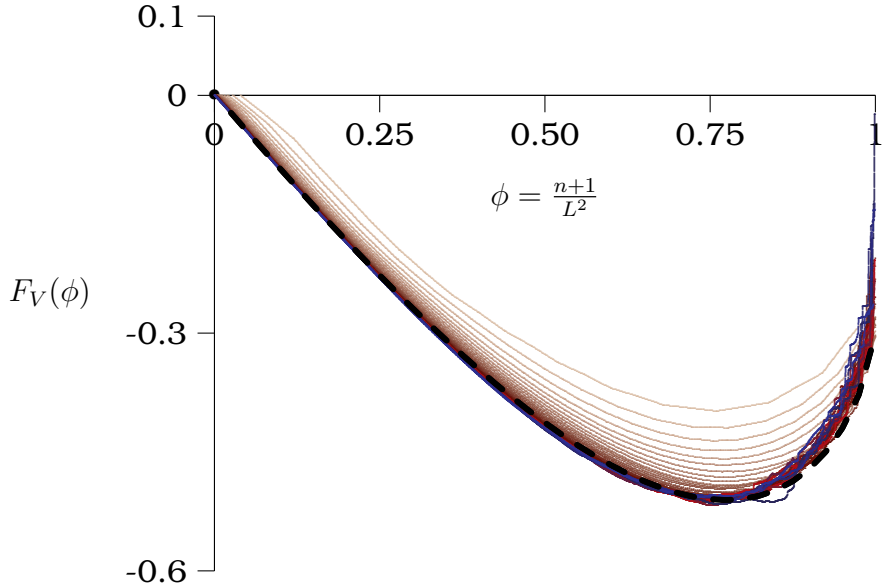

\input Figures/fig06.tex  
\caption{The estimated intensive free energy $F_V(\phi)$ per unit area 
obtained from flatPERM simulations for $5 \leq L\leq 40$.  These curves are plots 
of $F_V(\phi)$ as a function of $\phi$ and it appears that they converge to a limiting
curve with increasing $L$.  The curves increase in hue from tan ($L=5$) to 
blue ($L=40$).  In the dense regime (when $\phi \to 1^-$) the curves are more
noisy for larger values of $L$ due to numerical uncertainty at high concentration
(or for long compressed walks).  The dotted curve is the best fit to the data
by the mean field Flory-Huggins expression for $F_V(\phi)$ given in
equation \Ref{eqn27}.}
\label{fig06}  
\end{figure}

\section{Numerical determination of the osmotic pressure}

The fundamental quantity in the model in figure \ref{fig01} is $c_{n,L}$ (the number of compressed
self-avoiding walks of length $n$ in the square lattice;  see figure \ref{fig03}).
The extensive free energy is given by equation \Ref{eqn7} and the free energy
per unit area is $F_V(\phi) = - \sfrac{1}{L^2} \log c_{n,L}$ by equation
\Ref{eqn8A}, and since $V=L^2$.

We used the flatPERM-algorithm \cite{G97,PK04} to sample compressed walks
in $L\times L$ squares for $5\leq L \leq 40$.  For each value of $L\in\{5,6,7,\ldots,29\}$ 
the algorithm sampled walks of lengths $0 \leq n \leq L^2-1$ along 100,000 PERM sequences 
realised in the square.  For $L\in\{30,31,\ldots,40\}$ the sampling realised 200,000 PERM
sequences. 

PERM is an approximate enumeration algorithm, giving 
increasingly accurate estimates of $c_{n,L}$ the longer the algorithm runs.  
For $n$ approaching $L^2-1$ the algorithm sampled walks in the dense phase,
and while the algorithm did not succeed in sampling up to $n=L^2-1$ for 
large values of $L$,  it did approach this sufficiently close to give data for 
concentrations $\phi$ approaching $1$.  However, the quality of data, 
while good, deteriorated as $\phi \to 1^{-}$.  The free energy $F_V (\phi)$ was 
computed for data for $L=10$ and plotted in figure \ref{fig04} against
$\phi$ (these are the points denoted by {\Large$\bullet$}).

\begin{figure}[t]
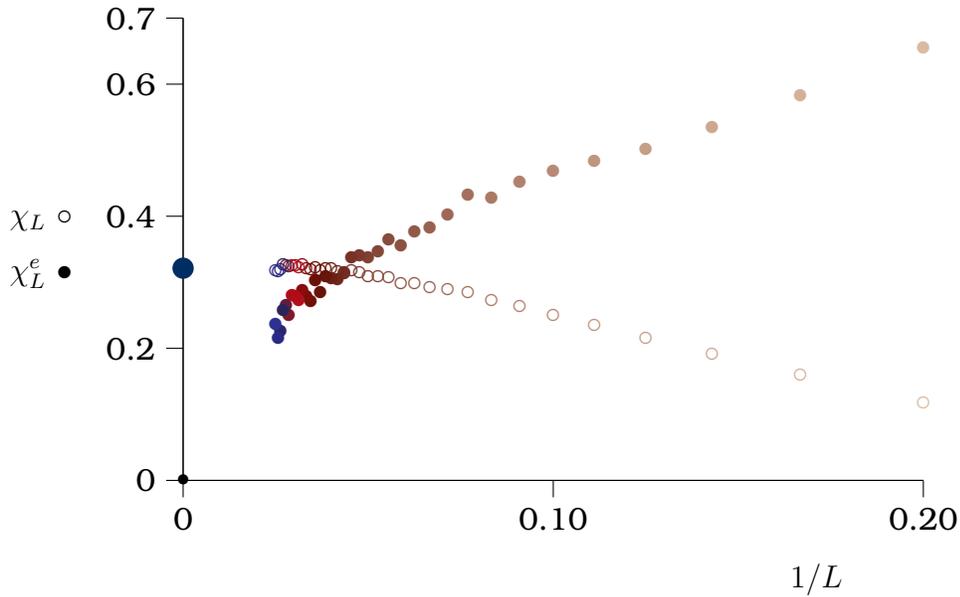

\input Figures/fig07.tex
\caption{The Flory interaction parameter $\chi$ as a function of $\frac{1}{L}$.
The solid bullets ($\bullet$) are the estimated effective values for $\chi$,
denoted by $\chi_L^e$, and determined in section \ref{section31}.  With
increasing $L$ these effective values decreases in magnitude, and does not
reach a stable value.  The open circles ($\circ$) are data for the estimated
values of $\chi$, denoted by $\chi_L$, and determined
in section \ref{section32} using the free energy $f_t(\phi)$.
For $L>20$ these estimates stabilise around a limiting value, and their
average gives the estimate $\chi_{saw} = 0.319 \pm 0.010$ for the
Flory interaction parameter of a compressed walk confined to a square in the
square lattice.  This value is indicated by the 
bullet ({\Large$\bullet$}) on the vertical axis.}
\label{fig07}  
\end{figure}

\subsection{The free energy $F_V(\phi)$}
\label{section31}

The Flory-Huggins mean-field expression for $F_V(\phi)$ is given 
by equation \Ref{eqn4} up to terms linear in $\phi$ (which were dropped 
from $E_{mix}$ in equation \Ref{eqn3}).  These linear terms are irrelevant 
in calculating $\Pi$ in equation \Ref{eqn14}, but are needed when the 
data in figure \ref{fig04} are analysed.  That is, we use the slight modification of 
the model in equation \Ref{eqn15A} to model the data in figure \ref{fig04}
by adding a linear term to the free energy with an unknown coefficient $a_0$.  That is,
\begin{equation}
F_V(\phi) \approx a_0\,\phi + \Sfrac{1}{L^2} \log \phi + (1-\phi)\log(1-\phi)
 - \chi_L\, \phi^2,
\label{eqn19A}   
\end{equation}
since $V=L^2$, and where we assume that the Flory interaction
parameter is $\chi_L$, a function of $L$.

Since the term in $\sfrac{1}{L^2}\log \phi$ diverges as $\phi\to 0^+$, it becomes dominant
as $\phi\to 0^+$.  Therefore, in fitting this expression to numerical data to determine
the parameters $a_0$ and $\chi_L$, we ignored data from low concentrations.
We did this consistently by leaving out data for values of $\phi < \sfrac{1}{L}$.
That is, the data for $\phi \in [\sfrac{1}{L},1]$ were used to determine
the parameters $a_0$ and $\chi_L$.   Examination of our results by first
leaving out the term $\sfrac{1}{L^2} \log \phi$, and then including it,
shows that the coefficients $a_0$ and $\chi_L$
are sensitive to this term.  Thus, we concluded that this
approach produces \textit{effective values} for these parameters, which we
denote by $a_0^e$ and $\chi_L^e$ in what follows.

The analysis for $L=10$ gives the following least squares fit to the data: 
\begin{equation}
\hspace{-1cm}
F_V(\phi)\vert_{L=10} 
\approx 0.1911\, \phi + 0.01 \log \phi
 +(1-\phi)\log(1-\phi) - 0.4676\, \phi^2 .
\label{eqn19}   
\end{equation}
This gives the effective values $a_o^e = 0.1911$ and $\chi_{10}^e = 0.4676$
for the parameters of the theory.  The model describes the data well, as shown by the 
continuous curve underlying the data points in figure \ref{fig04}.  

\begin{figure}[t]
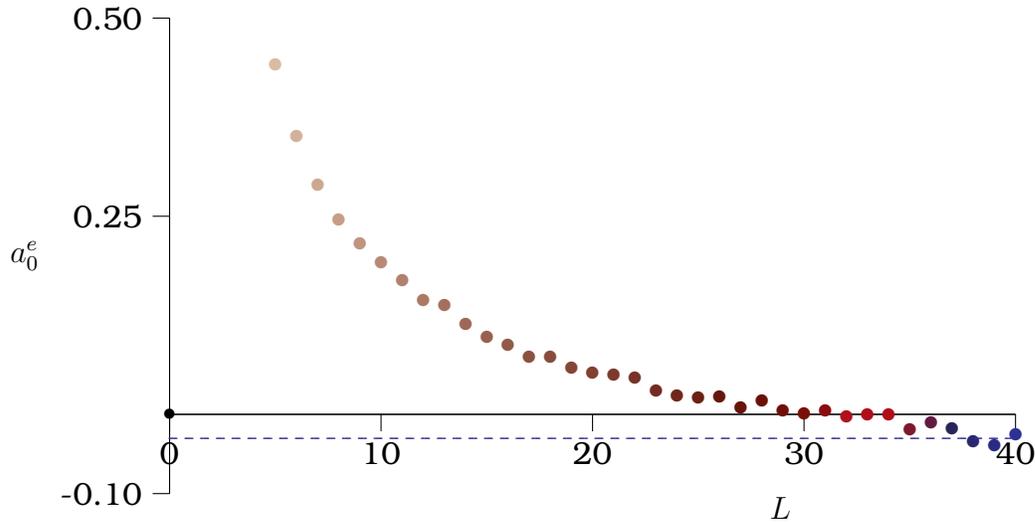

\input Figures/fig08.tex  
\caption{A plot of the least squares values of $a_0=(a_1+\chi_L)$ against $L$.   The data
points decrease with increasing $L$ and level off at the dotted line.}
\label{fig08}  
\end{figure}

The osmotic pressure for a compressed walk in an $L\times L$ square
can be directly computed from the above using equation \Ref{eqn13A}.  This shows that
\begin{equation}
\Pi\vert_{L=10} \approx 0.01(1-\log \phi) - \log(1-\phi) - \phi - 0.4676\,\phi^2 .
\end{equation}
This is plotted in figure \ref{fig05}.   Observe that this breaks
down as $\phi$ becomes very small.  The pressure should be increasing
with $\phi$, but due to the presense of the term $0.01 \log \phi$, $\Pi$
actually decreases for small values of $\phi$ -- this is unphysical, and is
also evidence that this term, at least for small values of $L$, causes
a deviation which does not capture the physical properties of the model.
Thus, in figure \ref{fig05} we plotted $\Pi$ only for $\phi>0.1
= \sfrac{1}{L}$.  The osmotic pressure $\Pi$ increases quickly with concentration when 
$\phi>0.3$.  As expected, it is also positive for all values of $\phi$.

The analysis leading to the effective value estimate $\chi_{10}^e \approx 0.4676$ 
for $L=10$ may be repeated for other values of $L$.  Since we collected 
data for $5\leq L \leq 40$, we estimated $\chi_L^e$ as a function of 
$\sfrac{1}{L}$ for $L\in\{5,6,7,\ldots,40\}$,  using the model above 
and for $\sfrac{1}{L} \leq \phi \leq 1$.   These results are plotted 
in figure \ref{fig07}.  We similarly plot the least squares estimates for $a_0^e$ 
in figure \ref{fig08}. For example,
\begin{equation}
\hspace{-2cm}
F_V(\phi) \vert_{L=40} 
\approx -0.02593\, \phi + 0.000625\log \phi
 +(1-\phi)\log(1-\phi) - 0.2363\, \phi^2 .
\label{eqn19}   
\end{equation}
This shows that the effective values at $L=40$ are $a_0^e = -0.02593$ and
$\chi_{40}^e = 0.2363$.

The plot of $\chi_L^e$ versus $\frac{1}{L}$ in figure \ref{fig07} 
(the data are shown by solid bullets $\bullet$) shows an accelerating 
decrease in the value of $\chi_L^e$ as $L$ increases.  A linear
least squares extrapolation of the data to the $y$-axis gives the effective
value at $L=\infty$ ($\chi_\infty^e = 0.21$) but not much significance
should be given to this result.

Similarly, the effective values of the coefficient $a_0^e$ are plotted against $L$
in figure \ref{fig08} and show a systematic decrease with increasing $L$. 
The estimates change sign and appear to converge to a negative constant as
$L\to\infty$.  A least squares fit to these data strongly suggests that
the limiting value is approximately ${-}0.03$, although not much significance can
be given to this as well.

\begin{figure}[t]
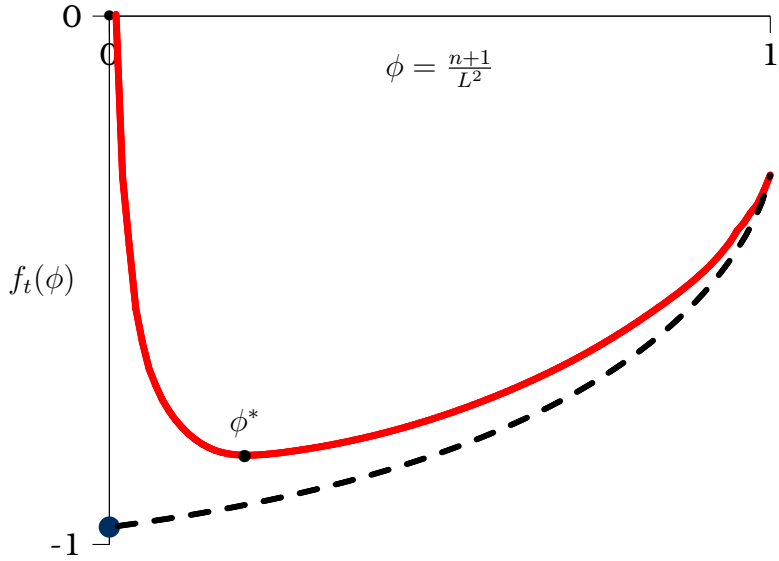

\input Figures/fig09.tex  
\caption{The free energy $f_t(\phi)$ for $L=10$.  It is a convex function
going through a minimum at the critical concentation $\phi^*$.  The 
dashed curve is the theoretical limiting free energy in the limit $L\to\infty$,
given by equation \Ref{eqn25} (and where we have assumed $\chi_{saw} = \frac{1}{3}$).  
With increasing $L$, it is expected that
$\phi^* \to 0$ and the minimum in $f_t(\phi)$ approaches the point
${-} \log \mu_2$ on the y-axis.}
\label{fig09}  
\end{figure}

\subsection{The free energy $f_t(b)$}
\label{section32}

\begin{figure}[t]
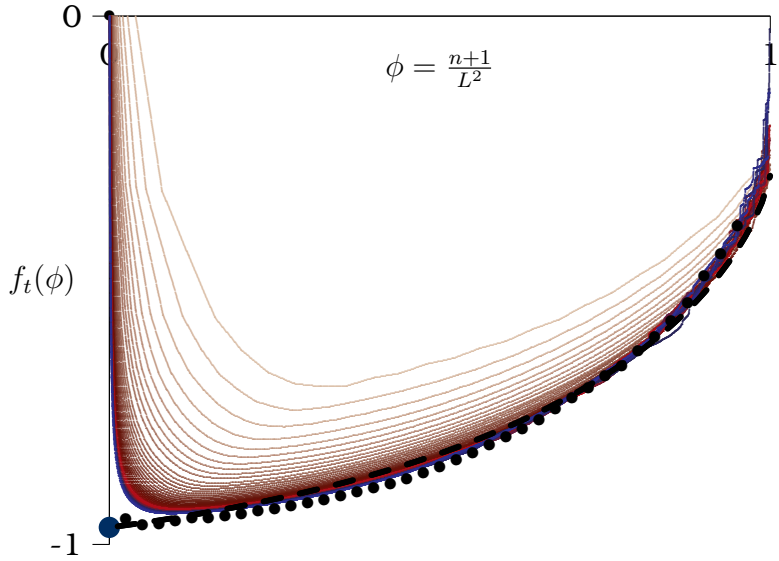

\input Figures/fig10.tex  
\caption{The free energy $f_t(\phi)$ as a function of $\phi$ for
$5 \leq L \leq 40$.  The curves decrease from $L=5$ to $L=40$ at the bottom.
Extrapolating the data using $c_0+c_1/V+c_2/V^{3/2}+c_3/V^2$ for fixed values of $\phi$
gives the bullets as $V\to\infty$.  The theoretical mean field limit is 
indicated by the dashed curve starting from ${-}\log\mu_2$ on the $y$-axis.  This 
approximation is good for $\phi$ not too small  and very close to the extrapolated 
values of the data.  In other words, the Flory-Huggins mean field approximation 
is not exact, although it is very successful in modelling the data.}
\label{fig10}  
\end{figure}

In order to estimate the Flory-Huggins parameter $\chi$ of the self-avoiding walk,
we turn our attention to the free energy per unit length, namely
$f_t(\phi)$ as defined in equation \Ref{eqn8B}.  In figure \ref{fig09} 
$f_t(\phi)$ is plotted for $L=10$ and it is a convex function with a minimum at
a critical concentration $\phi^*$. In figure \ref{fig10} these free energies
are plotted for $5\leq L \leq 40$.

With increasing $L$, $f_t(\phi)$ appears to approach a limiting curve, denoted by the
dashed line in figure \ref{fig09}.  Although the free energy is zero for
$\phi=0$,  it decreases quickly with increasing $\phi$ to the critical concentration
$\phi^*$ where $f_t(\phi)$ goes through its minimum value.  

For $\phi< \phi^*$ the walk is short in comparison with $L$ (this is the low concentration
regime), and the model is essentially a self-avoiding walk (in a wedge geometry) which grows 
as $\mu_2^{n+o(n)}$ with increasing length, where $\mu_2$ is the growth
constant of the square lattice self-avoiding walk \cite{BH57}.   That is, the number
of states of the system at low concentration is $O(\mu_2^{n+o(n)})$.

Moreover, as $L$ increases it is expected that $\phi^* \to 0$, since the walk can cross the square at
lower concentrations as it becomes compressed.  In the limit $L\to\infty$, 
$f_t(\phi)$ approaches $- \log \mu_2 \approx -0.970$ as $\phi\to 0^+$.  
This point is denoted by the bullet {\Large$\bullet$} on the vertical axis in figure \ref{fig09}.

\begin{figure}[t]
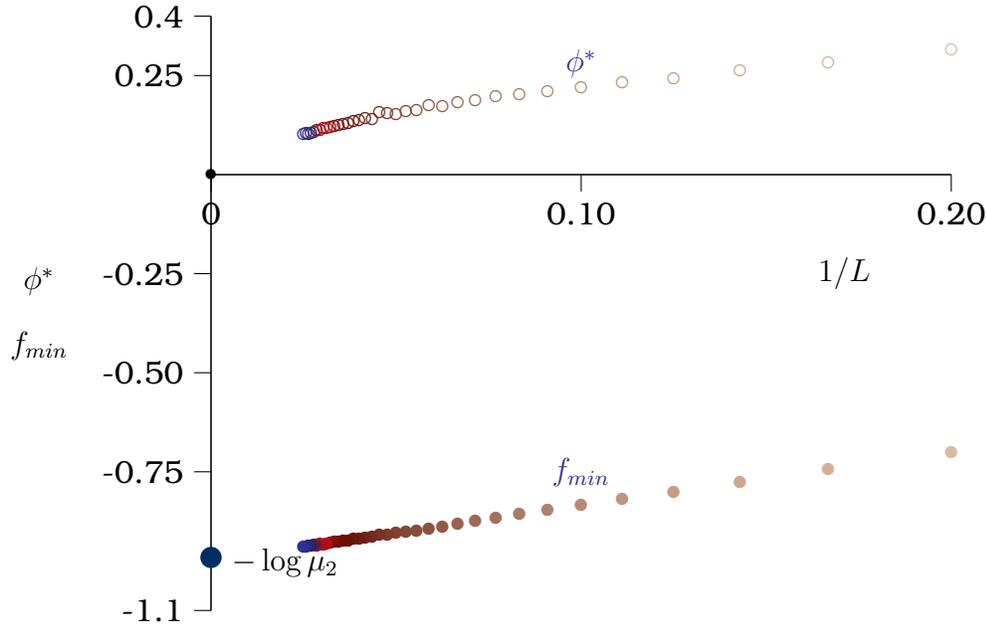

\input Figures/fig11.tex  
\caption{Extrapolating the critical concentration $\phi^*$ and the free
energy minimum $f_{min}$.  The solid bullets $\bullet$ are estimates
of the minimum free energy as a function of $\frac{1}{L}$.  A linear
extrapolation to the y-axis gives a value (denoted by the {\Large$\bullet$} on the
y-axis) very close to ${-}\log \mu_2$, as expected.  Similarly, the extrapolation
of $\phi^*$ gives a value very close to zero.}
\label{fig11}  
\end{figure}

Notice that the minimum in these curves
shifts to zero with increasing $L$ and that the minimum itself seems to approach
the value ${-}\log \mu_2$.  This is shown also in figure \ref{fig11}, where
we plot the critical concentration $\phi^*$ and the minimum of the free
energy, $f_{min}$, against $\sfrac{1}{L}$.  In the first case the data for
$\phi^*$ approach $0$, and the data for $f_{min}$ approach ${-}\log \mu_2$,
supporting the view above.  Linear extrapolation of the data gives ${-}0.972$, very
close to the numerical value of ${-}\log \mu_2$.

By equations \Ref{eqn8A} and \Ref{eqn8B}, $f_t(\phi) = \sfrac{1}{\phi} F_V(\phi)$.
Using the model in equation \Ref{eqn19A}, 
\begin{equation}
\hspace{-1cm}
f_t(\phi) 
= \Sfrac{1}{\phi}F_V(\phi) 
= a_0 + \Sfrac{1}{L^2\phi} \log \phi + \Sfrac{1-\phi}{\phi} \log(1-\phi) - \chi_L\,\phi .
\end{equation}
This gives a model for $f_t(\phi)$ as a function of concentration, which can be
used to determine the coefficient $a_0$ and the Flory interaction parameter
$\chi_L$.  Taking $L\to\infty$ gives the limiting curve
\begin{equation}
\lim_{L\to\infty} f_t(\phi) 
= a_0 + \Sfrac{1-\phi}{\phi} \log(1-\phi) - \chi_{saw}\, \phi,
\label{eqn24}   
\end{equation}
provided that $\phi>0$ and where $\chi_{saw}$ is the limiting value of the
Flory interaction parameter of the self-avoiding walk.  
Since $\chi_L$ approaches the value ${-} \log \mu_2$, and
the constant term in equation \Ref{eqn24} is $a_0-1$, this shows that
$a_0=1-\log \mu_2$ in the limit $L\to\infty$.  In this
model there is then only one parameter ($\chi_{saw}$) and we shall analyse
our data in order to estimate it.

Assuming that $\chi_{saw}$ is not large, the limiting free energy is increasing,
and so a fit to the finite $L$ estimates for $f_t(\phi)$ in figure \ref{fig09}
should not include the data on the descending side of $f_t(\phi)$ (that is,
for values of $\phi < \phi^*$).  In other words, we use the model in
equation \Ref{eqn24} and the data at concentrations $\phi$
with $\phi^* < \phi < 1$.  This gives the following model for finite $L$:
\begin{equation}
f_t(\phi) = 1-\log \mu_2 + \Sfrac{1-\phi}{\phi} \log(1-\phi) - \chi_L\, \phi, 
\qquad\hbox{for $\phi>\phi^*$}.
\label{eqn25}   
\end{equation}
For $L=10$ the data in figure \ref{fig09} gives $\chi_{10} = 0.2495$.

For the other values $L$ the results were similarly analysed and the
estimates for $\chi_L$ for $5\leq L \leq 40$ are shown by the points denoted
by open circles ($\circ$) in figure \ref{fig07}.  These estimates plateau 
for $L>20$, and if a simply average is taken for all the estimates of $L>20$, then
the limiting value is estimated to be
\begin{equation}
\chi_{saw} = 0.319 \pm 0.010
\label{eqn26}   
\end{equation}
with a stated $95$\% error bar and which we round to $\chi_{saw} = 0.32(1)$.   
Doing a linear extrapolation of the data
with $L>20$ instead gives $\chi_{saw}  = 0.331\ldots$, just slightly outside the 
confidence interval above.  Thus, we take as our best estimate for 
the Flory interaction parameter of compressed square 
lattice self-avoiding walks the result in equation \Ref{eqn26}, but also
note that this does not rule out a slightly larger value. 

\section{Conclusions}

In this paper we examined a compressed self-avoiding walk in the mean
field by shoehorning it into Flory-Huggins theory.  We showed that 
Flory-Huggins theory is very successfull in modelling the free energies 
$F_V(\phi)$ and $f_t(\phi)$; in the first case by effective values 
for the Flory interaction parameter, and in the second case 
by computing a consistent value of this parameter.  Only a 
minor modification (the addition of an explicit linear term in the 
Flory-Huggins mean field free energy) was needed to achieve this.

The estimate of the Flory interaction parameter $\chi_{saw}$ for
the self-avoiding walk was
best done by using the model in equation \Ref{eqn25} and this produced
the finite $L$ values denoted by open circles ($\circ$) in figure \ref{fig07}.  
Since these estimates quickly level off for $L>20$, a simple average was
taken to obtain the best estimate for $\chi_{saw}$ in equation 
\Ref{eqn26}.  This also shows that the Edwards excluded volume parameter
of two dimensional self-avoiding walks in the square lattice is
$\upsilon = 1-2\chi_{saw} = 0.362\pm 0.020$.

The best estimate for $\chi_{saw}$  does not rule out the possibility that
$\chi_{saw} = \sfrac{1}{3}$ (in which case $\upsilon = \sfrac{1}{3}$).  
In fact, this value gives very good fits
to our data as shown by the dashed curves in figures \ref{fig06} and
\ref{fig10}.  That is, the limiting free energy of compressed square 
lattice walks at concentration $\phi$ is approximated well by
\begin{equation}
\lim_{L\to\infty} f_t(\phi)
= \C{F}_t (\phi)  \approx 1-\log \mu_2+ \Sfrac{1-\phi}{\phi} \log(1-\phi)
 - \Sfrac{1}{3}\, \phi.
\label{eqn25}   
\end{equation}
Taking $\phi \to 0^+$ gives the limiting free energy of self-avoiding
walks:  $\C{F}_t(0) = - \log \mu_2$ so that the number of self-avoiding
walks grows as $\mu_2^{n+o(n)}$, as expected.  

The case $\phi=1$ is a model of Hamiltonian walks in a square of
side-length $L$.   The limiting free energy per unit area (or per
unit length) is approximately given by $\C{F}_t(1) \approx \sfrac{2}{3} - \log \mu_2$,
assuming that $\chi_{saw} = \sfrac{1}{3}$.  
This shows that the number of Hamiltonian walks grows approximately 
at the rate $(\mu_2\,e^{-2/3})^{L^2+o(L^2)}$, or the connective 
constant of Hamiltonian walks is approximately given by
\begin{equation}
\kappa_H \approx \log \mu_2 - \sfrac{2}{3} = 0.3034\ldots.
\end{equation}
This approximation is, of course, a result of the Flory-Huggins
expressions for the free energy, and one should recall in this context
that third and higher order interactions are ignored in those
mean field expressions.

The limiting free energy per unit area should also be given by
\begin{equation}
\lim_{L\to\infty} F_V(\phi)
= (1-\log \mu_2)\,\phi + (1-\phi)\log(1-\phi) - \chi_{saw}\, \phi^2.
\label{eqn27}   
\end{equation}
This curve is plotted as the dashed curve in figure \ref{fig06},
and it describes the data well (with $\chi_{saw} = \sfrac{1}{3}$).

The osmotic pressure in the thermodynamic limit can now 
be computed using equation \Ref{eqn10}, and is given by
\begin{equation}
\Pi_{saw} = - \log(1-\phi) - \phi - \chi_{saw} \phi^2.
\label{eqn28}  
\end{equation}
Series expansion of this shows that
$\Pi_{saw} = (\sfrac{1}{2} - \chi_{saw})\, \phi^2 + \sfrac{1}{3}\, \phi^3 + \ldots$
for small values of $\phi$.  This shows that $\Pi_{saw}|_{\phi=0} = 0$ and 
$\Pi_{saw}>0$ and increasing for $\phi>0$.  
If $\chi_{saw}=\sfrac{1}{3}$, then the osmotic pressure for small concentrations is given by
$\Pi_{saw} \approx \sfrac{1}{6}\, \phi^2 + \sfrac{1}{3}\, \phi^3 + \ldots$;
see, for example, equation III.14 in reference \cite{deG79} as well.
For $\phi\to 1^-$ the osmotic pressure diverges logarithmically, as seen
in equation \Ref{eqn28}.

We are continuing our examination of the Flory-Huggins description
of dense systems of self-avoiding walks, including models of
interacting self-avoiding walks, and models in $d=3$.

\vspace{0.5cm}
\noindent{\bf Acknowledgements:} EJJvR acknowledges financial support 
from NSERC (Canada) in the form of a Discovery Grant.   We are grateful to
SG Whittington for helpful remarks and advice.

\vspace{1cm}
\noindent{\bf References}
\bibliographystyle{plain}
\bibliography{References}

\end{document}